\documentstyle[epsf,12pt]{article}

\begin{document}
\begin{titlepage}
\centerline{\bf{PHASE TRANSITION AND HYBRID STAR }}
\centerline{\bf{IN A SU(2) CHIRAL SIGMA MODEL}}
\vspace{0.5in}
\centerline { P. K. Jena$^*$ and L. P. Singh}
\date{}
\vspace{0.25in}
\centerline{Department of Physics,Utkal University,Vanivihar,}
\centerline{ Bhubaneswar-751004,India.}
\vspace{1in}
\centerline{\bf{Abstract}}
\vspace{.1in}

  We use a modified SU(2) chiral sigma model to study nuclear matter at high 
density using mean field approach. We also study the phase transition
of  nuclear matter to quark matter in the interior of highly dense neutron
stars. Stable solutions of Tolman-Oppenheimer-Volkoff equations representing
hybrid stars are obtained with a maximum mass of 1.69$M_{\odot}$, 
radii around 9.3 kms  and  a quark matter core constituting nearly 
55-85$\%$ of the star radii.

\vspace{0.5in}


\vspace{2.0in}
\underline{$^*$email: pkjena@iopb.res.in}

\end{titlepage}

\section{Introduction}

It has been argued that nuclear matter undergoes a phase transition to 
quark matter at  high densities and/or high temperatures. It is expected that  
the high temperature limit  has interesting consequences in heavy ion       
collision and/or in cosmology, but  high baryon density behavior is 
important to the study of neutron stars .

  The quark structure of hadrons implies that at sufficiently large
nuclear densities nuclear matter should convert into quark matter. 
The density at
which transition occurs, is believed to be a few times nuclear matter density .
The lattice calculations indicate that for nonzero quark masses the phase 
transition may be weak first order or second order$^{1}$. Most of the model 
calculations find it to be first order. Thus for large enough mass of neutron 
star, its core may consists of quark matter. In addition, if the phase
transition is first order, a part of the core may consists of mixed phase of 
quark and nuclear matter. Kapusta and others$^{2,3}$  have used a
nonlinear Walecka model for the nuclear phase and a bag model for the
quark phase . They have found that if the hadron-quark transition
density $n_B \geq 4n_0 $($n_0$ is the nuclear matter density
=0.153fm$^{-3}$),then it is quite unlikely that stable stars with quark 
interior exist. Ellis $\it {et \ al.}$$^3$ have also studied  the
possibility of a second order hadron-quark phase transition. In such
cases, one needs to include an additional phenomenological parameter. 
A number of studies using different models have also been
undertaken$^{4-7}$ showing a first order phase transition.

  In all the above works, $ \mu_B$ (baryon chemical potential)is the 
only conserved charge and pressure remains constant in the mixed phase. 
Glendenning $\it{ et \ al.}$$^8$ and  Burgio $\it{ et \ al.}$$^9$  
have considered both $\mu_B $ and $\mu_E$ (electron chemical 
potential ) as conserved charges. This has the consequence that
pressure varies continuously for all mixtures of the two
components(hadron and quark) throughout the mixed phase.

  In the present work, we have used a modified SU(2) chiral sigma
model(MCH)$^{10}$ for hadronic matter since chiral model has been
very  successful, as such, in describing high density nuclear matter. 
The importance of chiral symmetry$^{11}$ in the study of nuclear matter 
was first emphasized by Lee and Wick$^{12}$  and has become over the years, 
one of most useful tools to study high density nuclear matter 
at the microscopic level.   
The nonlinear terms in the chiral sigma model give rise to the three-body 
forces which become significant in the high density regime$^{13}$. Further, 
the energy per nucleon at saturation needed the introduction of
isoscalar field$^{14}$  in addition to the scalar field of
pions$^{15}$. We also include the interaction due to isospin triplet 
$\rho$-vector meson to describe the neutron rich-matter$^{16}$.

   The modified SU(2) chiral sigma model$^{10}$ considered by us   
includes two extra higher order 
scalar field interaction terms which ensures an appropriate
incompressibility of symmetric nuclear matter at saturation
density. Further, the equation of state(EOS) derived from this model
is compatible with that inferred from recent heavy-ion collision data$^{17}$.

  In our work, we consider the baryon chemical potential $\mu_B$ as the 
only conserved charge. Consequently  pressure remains constant in the mixed 
phase region. A first order phase transition between beta stable
nuclear matter and quark matter is indicated.  
Taking the existence of such a phase transition between nuclear
matter and quark matter as a guide, we solve the
Tolman-Oppenheimer-Volkoff(TOV) equations with appropriate nuclear
matter$^{10}$ and quark matter$^{18}$ equations of state and find the
hybrid stars to consist of a quark-matter core with the nuclear matter 
forming the crust.

  This paper is organised as follows. In sec.2, we present the equation
of state for nuclear matter. The quark matter equation of state is 
discussed in sec.3. In sec.4, we discuss the structure of hybrid star.  
We discuss and summarise our results in sec.5.
 
\section{\bf{Nuclear matter Equation of State }}

   The modified SU(2) chiral sigma model considered by us is
described by the  Lagrangian density$^{10}$,
\begin{eqnarray}
 \it{L} = \frac{1}{2}(\partial_{\mu} \vec{\pi}.\partial^{\mu}\vec{\pi} + 
   \partial_{\mu}\sigma  
  \partial^{\mu}\sigma )-\frac{1}{4}F_{\mu \nu} F_{\mu \nu}-
   \frac{\lambda}{4}(x^2-x_0^2)^2 
  -\frac{\lambda B}{6m^2}(x^2-x_0^2)^3 \nonumber \\ 
   -\frac{\lambda C}{8m^4}(x^2-x_0^2)^4 - g_{\sigma}\bar{\psi }(\sigma +
  i\gamma_{5}\vec{\tau} .\vec{\pi} )\psi  
 +\bar{\psi}(i \gamma_{\mu}
  \partial ^{\mu} -g_{\omega}\gamma_{\mu}\omega ^{\mu})\psi \nonumber \\   
   +\frac{1}{2}g_{\omega}^2  x^2 \omega_{\mu}\omega ^{\mu} 
  -\frac{1}{4} G_{\mu \nu}.G^{\mu \nu}+\frac{1}{2}
   m_{\rho}^{2}\vec{\rho_{\mu}}.\vec{\rho^{\mu}}
   -\frac{1}{2}g_{\rho}\bar{\psi}(\vec{\rho_{\mu}}.\vec{\tau}\gamma^{\mu})\psi 
  \end{eqnarray}
\noindent
In the above Lagrangian, $F_{\mu \nu} \equiv \partial_{\mu}\omega_{\nu} 
 - \partial_{\nu} \omega_{\mu}$, $G_ {\mu \nu} \equiv \partial_{\mu}\rho_{\nu} 
 - \partial_{\nu} \rho_{\mu}$ and $x = (\vec{\pi}^2 +\sigma ^2)^{1/2}$, 
$\psi $ is the nucleon  isospin doublet, $\vec{\pi}$ is the 
pseudoscalar-isovector pion field, $\sigma$ is the scalar field and 
$\omega_{\mu}$, is a dynamically generated isoscalar vector field, which
couples to the conserved baryonic current
$j_{\mu}=\bar{\psi}\gamma_{\mu}\psi$. $\vec{\rho_{\mu}}$ is the
isotriplet vector meson field with mass $m_{\rho}$. B and C are
constant  coefficients 
associated  with the higher order self-interactions of the scalar field .

   The masses of the nucleon, the  scalar meson  and the vector meson
are respectively given by 

\begin{equation}
  m = g_{\sigma}x_0, \  m_{\sigma} =\sqrt{2\lambda} x_0, \ 
 m_{\omega} = g_{\omega}x_0 
\end{equation}
\noindent
 Here $x_0$ is the vacuum expectation value of the  $\sigma $ field , 
$ g_{\omega}$, $g_{\rho}$ and $g_{\sigma}$
are the coupling constants for the vector and scalar fields respectively  
 and $\lambda =
(m_{\sigma}^2 - m_{\pi}^2)/ (2 \it {f}_{\pi}^2) $, where  $m_{\pi}$ is the 
pion mass , $\it{f}_{\pi}$  is the pion decay coupling constant .

 Mean-field approximation has been used extensively to obtain field theoretical
equation of state  for high density matter$^{19}$. Using this
approximation,  the equation of motion for isoscalar vector field is 
\begin{equation}
   \omega_{0}=\frac{n_{B}}{g_{\omega}x^2}
\end{equation}
\noindent 
 and the equation of motion for the scalar field in terms of  
$ y\equiv \frac{x}{x_0}$ is of the form$^{10}$

\begin{equation}
 (1-y^2)-\frac {B}{m^2C_{\omega}}(1-y^2)^2 +\frac {C}{m^4C_{\omega}^2}(1-y^2)^3 
    +\frac{2 C_{\sigma}C_{\omega}n_{B}^2}{m^2y^4}-\frac{C_{\sigma}\gamma}{\pi^2} 
     \int_{0}^{k_f} \frac{k^2 dk}{\sqrt{k^2+{m^*}^2}} = 0
\end{equation}
\noindent
where $m^* \equiv ym$ is the effective mass of the nucleon and the coupling 
constants are 

\begin{equation}
   C_{\sigma} \equiv  \frac {g_{\sigma}^2}{m_{\sigma}^2},\ \ 
   C_{\omega}\equiv \frac {g_{\omega}^2}{m_{\omega}^2} \nonumber
\end{equation} \nonumber
\noindent
  
  The baryon number density $n_B = n_p + n_n = \frac{\gamma
k_f^3}{6\pi^2} $ , where $k_f$  
is the Fermi momentum and $\gamma$ is the spin degeneracy factor which is 
equal to 4 and 2 for nuclear and neutron matter respectively.
The equation of motion for $\rho $ , in the mean field approximation
gives$^{15}$
\begin {equation}
 \rho_{0}^{3} = (g_{\rho}/2m_{\rho}^{2})(n_p - n_n)
\end{equation} 

  At high densities the interior of neutron stars composed of
asymmetric nuclear matter with an admixture of electrons. The
concentrations of neutrons,protons and electrons can be determined using
conditions of beta equilibrium and electrical charge neutrality$^{20}$.

 \ \ \ \ \ $ \mu_{n} = \mu_{p}+\mu_{e}$ ; $n_p = n_e $ , 

\noindent
(here $\mu_i $ is the chemical potential of the particle species
$\it{i}$). In our analysis we find the ratio $n_p/n_n$ to lie in the
range 0.003-0.08 for $n_B$ taking value from 0.02 fm$^{-3}$ to 0.25
fm$^{-3}$. We have included the interaction due to isospin triplet
$\rho$-meson in Eqn-1 for describing neutron-rich matter.  The
symmetric energy coefficient that follows from the semi-empirical
nuclear mass formula is$^{10,15}$ \\

\ \ \ \ \  $ a_{sym} = \frac{C_{\rho}k_{f}^{3}}{12 \pi^{2}}
   +\frac{k_f^{2}}{6 \sqrt{k_f^2 +{m^*}^2}}$ , 

\noindent
where $C_{\rho} \equiv g_{\rho}^2/m_{\rho}^2$ .

   The diagonal components of the conserved total stress tensor corresponding 
to the Lagrangian(Eqn.1) together with the mean field equation of motion 
for the fermion field and a mean-field approximation for the meson fields is 
used to calculate the equation of state . The total energy
density($\epsilon $)  and pressure(P), for the neutron rich nuclear
matter in $\beta$-equilibrium is given by$^{10}$ 

\begin{eqnarray}
   \epsilon= \frac{m^2(1-y^2)^2}{8C_{\sigma}}-\frac{B}{12C_{\omega} C_{\sigma}}
 (1-y^2)^3 +\frac{C}{16m^2C_{\omega}^2C_{\sigma}}(1-y^2)^4 \nonumber\\
+\frac{C_{\omega}n_B^2}{2y^2}+
   \frac{\gamma}{2\pi^2} \sum_{n,p,e}\int_{0}^{k_f} k^2dk
   \sqrt{k^2+{m^*}^2} + \frac{1}{2}m_{\rho}^2 (\rho_{0}^3)^2 ,
\nonumber \\ 
  P= -\frac{m^2(1-y^2)^2}{8C_{\sigma}} +\frac {B}{12C_{\omega} C_{\sigma}}
   (1-y^2)^3-\frac{C}{16m^2C_{\omega}^2C_{\sigma}}(1-y^2)^4 \nonumber\\
 +\frac{C_{\omega}n_B^2}{2y^2}+
  \frac{\gamma}{6\pi^2}\sum_{n,p,e}\int_{0}^{k_f}
   \frac{k^4dk}{\sqrt{k^2+{m^*}^2}}+ \frac{1}{2}m_{\rho}^2 (\rho_{0}^3)^2
\end{eqnarray}
\noindent
The energy per nucleon is $\frac {E}{A}=\frac {\epsilon}{n_B}$ and the
chemical potential is 
 $\mu=(P+\epsilon)/n_B $. 

   The values of five parameters $C_\sigma, C_{\omega},C_{\rho} $, B and
C occurring
in the above equations are obtained by fitting with the saturation values of 
binding energy/nucleon 
(-16.3 MeV), the saturation density (0.153 fm$^{-3}$), the symmetric
energy(32 MeV), the effective(Landau) mass
(0.85M)$^{21}$, and nuclear incompressibility ($\sim $300 MeV), in accordance
with recent heavy-ion  collision data$^{17}$ are   
 $C_{\omega}$ = 1.999 fm$^2$, $C_{\sigma}$ = 6.8157
fm$^2$, $C_{\rho}$ = 4.661 fm$^2$, B = -99.985 and C = -132.2456.
\begin{figure}[t]
\leavevmode
\protect\centerline{\epsfxsize=5in\epsfysize=4.5in\epsfbox{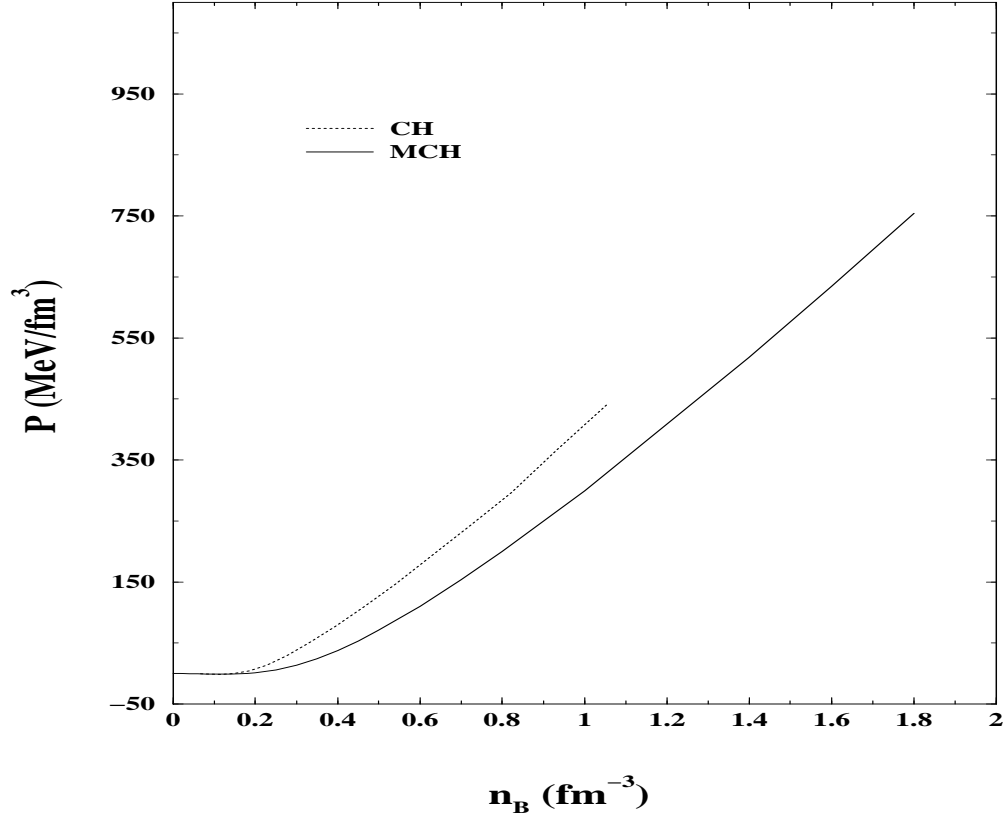}}
\caption{Pressure (P) vs. baryon  number density($n_B$) for nuclear matter } 
\end{figure}
 
   The pressure as a function of number density of nuclear matter is
presented in Fig.[1]. The solid curve(MCH) corresponds to the model
considered by us, whereas the dotted curve(CH) corresponding to the
original chiral sigma model$^{15}$ is presented for comparison. 
The equation of state for the present model is found to be softer 
with respect to the original  one.


\section{\bf{Quark Matter Equation of State and Phase Transition.}}

  Several authors$^{22-25}$ have studied the possible existence of
quark matter in the core of neutron stars/pulsars. Densities of these 
stars are expected to be high
enough to force the  hadron  constituents or nucleons to strongly overlap
thereby yielding quark  matter. Since  the distance involved
is small, perturbative  Quantum Chromodynamics (QCD)  is used to derive quark
matter equation  of state . We consider here the quark matter EOS which
includes u, d and s quark degrees of freedom$^{18,25}$ in addition to
electrons.  We have taken the electron, up and down quark masses to be   
zero$^{18}$ and  the strange quark mass is taken to be 180 MeV. In 
chemical equilibrium, one has $\mu_d$= $\mu_s=\mu_u+\mu_e $ which can be 
written in terms of baryon and electric charge chemical potentials as$^{5,6}$
\begin{eqnarray}
\mu_u = \frac{1}{3}\mu_B+\frac{2}{3}\mu_E, \nonumber\\
\mu_d = \frac{1}{3}\mu_B-\frac{1}{3}\mu_E,\nonumber \\
\mu_s = \frac{1}{3}\mu_B-\frac{1}{3}\mu_E \nonumber \\ \noindent
\mu_e = -\mu_E
\end{eqnarray}
\noindent
The pressure contributed by the quarks is computed to order $\alpha =\frac{g^2}
{4\pi}$, where g is the QCD coupling constant.  The electron pressure
is$^{18}$ 

\begin{equation}
 P_e = \frac{\mu_e^4}{12\pi^2} \ ,
\end{equation}
\noindent
and the pressure for quark flavor $\it {f}$, with $\it {f}$ = u, d or s is
$^{3,18,25}$

\begin{eqnarray}
 P_{f} = \frac{1}{4\pi^2}[\mu_f k_f(\mu_f^2-2.5m_f^2)+1.5m_f^4 ln (\frac{\mu_f+  
  k_f}{m_f})]\nonumber\\
-\frac{\alpha_{s}}{\pi^3}[\frac{3}{2}(\mu_f k_f-m_f^2 
  ln (\frac{\mu_f+k_f}{m_f}))^2 - k_f^4]
\end{eqnarray}

\noindent
 Where  $k_f=(\mu_f^2-m_f^2)^{1/2}$ is the Fermi momentum. The total
pressure, including   the bag constant B is given by$^{18}$ 
\begin{equation}
 P=P_e+\sum_{f} P_f -B
\end{equation}
\noindent
In the above equations $\mu_B$ and $\mu_E$ are only two independent 
chemical potentials. $\mu_E$ is adjusted so that the matter is
electrically neutral i.e $ {\partial P}/{\partial \mu_B} =0 $ . 
The baryon number density ($n_B$) and the energy density ($\epsilon$)
for quark matter can be derived using the thermodynamic
relations$^{18}$ 

\ \ \ \ \ \ \ \ \ \ \ \ \ \ \ \ \ \ \ \ \ $n_B = \partial P/\partial \mu_B $  
\ and \  $\epsilon = -P+\mu \frac{\partial P}{\partial \mu}$ .

 Now  we shall study the possible scenario of phase transition from nuclear
matter to quark matter . Gibb's criteria is used to determine the phase 
boundary of the co-existence region between the nuclear and quark phase.  
The critical pressure and the critical
chemical potentials  are determined by the condition 
\begin{equation}
  P_{nm}(\mu_B)=P_{qm}(\mu_B)
\end{equation}
\noindent
We have taken typical values $\alpha_s$ = 0.5, 0.6 and the bag constant  
B = (155 MeV)$^4$ and  (150 MeV)$^4$, which are reasonable values to 
calculate pressure  in the quark sector$^{5}$.

\begin{figure}[t]
\leavevmode
\protect\centerline{\epsfxsize=5in\epsfysize=5in\epsfbox{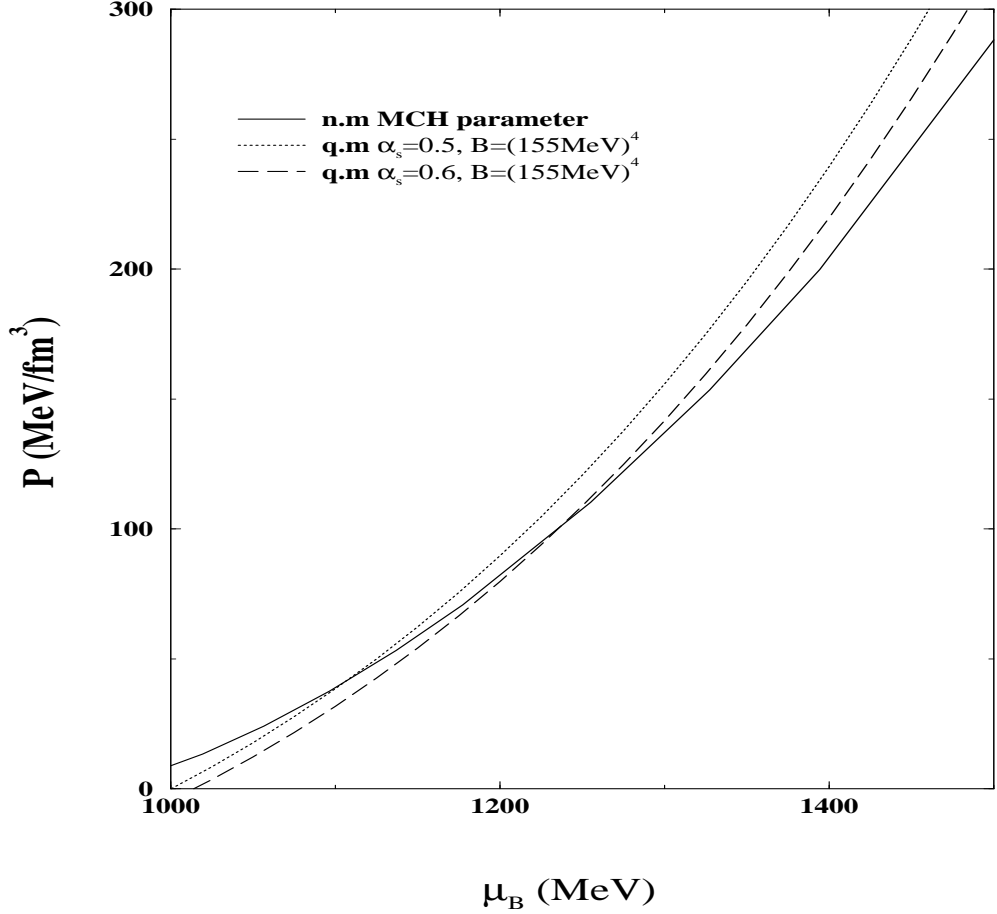}}
\caption{Pressure (P) vs. chemical potential for nuclear  matter and
  quark matter with various $\alpha_s$ at constant bag pressure (B)} 
\end{figure}

\begin{figure}[here]
\leavevmode
\protect\centerline{\epsfxsize=5in\epsfysize=5in\epsfbox{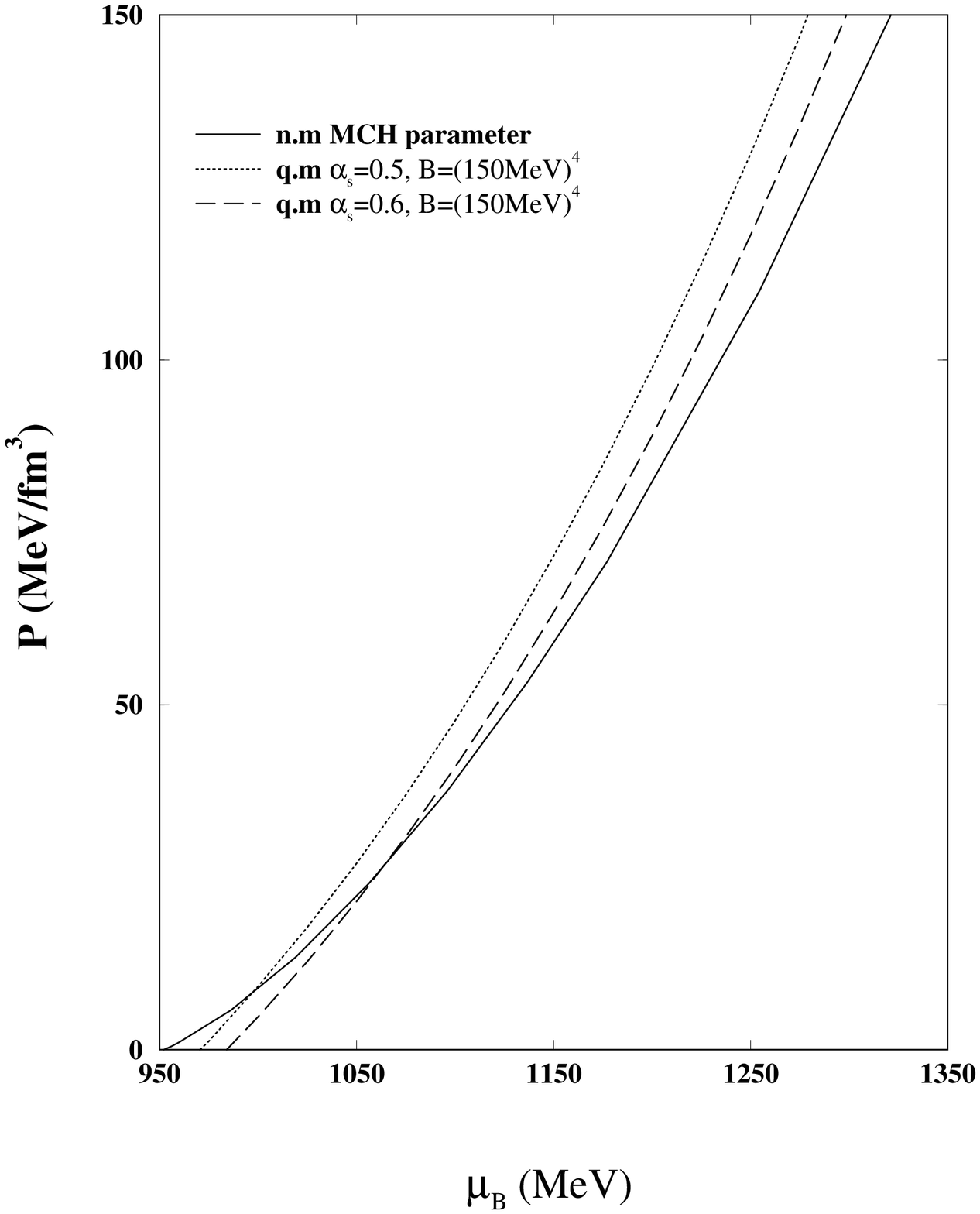}}
\caption{Pressure (P) vs. baryon chemical potential for nuclear matter and
  quark matter with various $\alpha_s$ with different bag pressure (B)} 
\end{figure}

     We have plotted pressure versus chemical potential for beta
stable nuclear matter and quark matter in Fig.[2]. The solid line is
shown for the beta-stable nuclear matter with MCH parameter set. The dotted and
dashed lines correspond to quark matter with $\alpha_s = 0.5$ and $\alpha_s =
0.6$ with bag pressure  B = (155 MeV)$^4$ respectively .  We find that 
there exist phase transition points for nuclear matter at different 
pressures and chemical potentials. The transition point $(P_{crit.}, 
\mu_{crit.})$ for $\alpha_s $ = 0.6  is (101 MeV/fm$^3$, 1237 MeV)
and that for $\alpha_s =  0.5$ is (43 MeV/fm$^3$, 1111 MeV).
Fig.[3] shows the phase  diagram with a different bag pressure 
$B = (150MeV)^4$. Here the transition point ($P_{crit.}$, $\mu_{crit.}$) for
$\alpha_s$ = 0.6 is  (27 MeV/fm$^3$, 1064 MeV) and that for
$\alpha_s $ = 0.5 is (7 MeV/fm$^3$, 993 MeV) . Thus, we find though
the transition point increases with the increase of bag pressure and
$\alpha_s$, with one of them remaining constant, the dependence on
$\alpha_s$ is more sensitive.

  Now considering a typical transition point ($P_{crit.}$, $\mu_{crit.}$)
for B=(150MeV)$^4$, $\alpha_s$= 0.5, at the critical pressure, the energy 
densities for the quark matter $(\epsilon_{crit.}^{qm})$ and nuclear 
matter $(\epsilon_{crit.}^{nm})$ sectors  are found to be  387  
MeV/fm$^3$ and 252 MeV/fm$^3$ respectively. The baryon number densities
corresponding to the critical $\mu_B$ in quark matter is $n_{B}^{qm} = 
0.37 fm^{-3}$ and that in nuclear matter  is $ n_{B}^{nm}=0.27
fm^{-3}$ which is reasonably an order of magnitude (about 1.8 to 2.5 times)
higher than the nuclear matter density. The discontinuity in the
number density as well as energy density indicates a first order
phase  transition. This phase transition from nuclear matter to quark 
matter obviously implies that the interior of neutron star consists of 
quark matter. We investigate this possibility further in the next section.


\section{\bf{Hybrid Stars.}}

   Having established the existence of a phase transition, we now proceed to 
study the structure of a hybrid star. For the description of neutron star   
which can generate curvature in the space time geometry due to high 
concentration of matter, one has to apply Einstein's general theory of 
relativity. The space-time geometry generated by a spherical neutron star, 
described by the Schwarzschild metric can be represented in the form$^{26}$
\begin{equation}
 ds^2= -e^{\nu(r)}dt^2 +[1-2M(r)/r]^{-1} dr^2+ r^2[d\Theta ^2+sin^2\Theta 
  d\phi^2].
\end{equation}
\noindent
The Tolman-Oppenheimer-Volkoff (TOV) equations which determine the 
star structure and the geometry, in  dimensionless
forms$^{26,27}$, are given by 

\begin{equation}
 \frac{d\hat{P}(\hat{r}r_0)}{d\hat{r}}=-\hat{G}\frac {[\hat{\epsilon}
  (\hat{r}r_0)+\hat{P}(\hat{r}r_0)][\hat{M}(\hat{r}r_0)+4\pi a\hat{r}^3
   \hat{P}(\hat{r}r_0)]}
      {\hat{r}^2[1-2\hat{G}\hat{M}(\hat{r}r_0)/\hat{r}]} \ ,  
\end{equation}

\begin{equation}
 \hat{M}(\hat{r}r_0)=4\pi a
 \int_0^{\hat{r}r_0}d\hat{r}^{\prime} \hat{r^\prime}^{2}
    \hat{\epsilon}(\hat{r}^{\prime}r_0) \ ,  
\end{equation} 
\noindent
and the metric function, $\nu (r)$, relating the element of time at $r=\infty$
 is given by$^{26}$ 

\begin{equation} 
 \frac{d\nu (\hat{r}r_0)}{d\hat{r}}=2\hat{G}\frac{[\hat{M}(\hat{r}r_0)
   +4\pi a\hat{r}^3\hat{P}(\hat{r}r_0)]}
      {\hat{r}^2[1-2\hat{G}\hat{M}(\hat{r}r_0)/\hat{r}]} \ . 
\end{equation}
\noindent
 The following substitutions  have been made in above Eqns.(14-16).
\begin{equation} 
 \hat{\epsilon} \equiv \epsilon /\epsilon_c ,\ \ \hat{P}\equiv P/\epsilon_c ,
 \ \  \hat{r}\equiv r/r_0 ,\ \  \hat{M}\equiv M/M_{\odot} 
\end{equation}
\noindent
Here , 
\begin{equation}
 a \equiv \epsilon_c r_0^3/M_{\odot}, \ \ \hat{G}\equiv \frac
 {GM_{\odot}}{f_1r_0}
\end{equation}
\noindent
with $f_1$ = 197.327 MeV fm and $r_0 = 3\times$ 10$^{19}$ fm.\\
The quantities with hats are dimensionless in  above equations. The 
gravitational constant  G = 6.7079 $\times$ 10$^{-45}$ MeV$^{-2}$.

      For complete calculation of a stellar  model, one has to integrate  
Eqns.(14-16) from the star's center at r = 0 with a given central density 
$\epsilon_c$ as input until the pressure P(r) at the surface vanishes.
With any reasonable central energy density, we expect that at the center 
we shall have only quark matter. Hence we shall be using
here the equation of state for quark matter through Eqn.(10) with
$\hat{P}(0) = P(\epsilon_c)$. We then integrate the TOV equations until the 
pressure and density decrease to their critical values at $r = r_c$.
For $r > r_c$, we shall have equation of state for $\beta$-stable nuclear
matter where pressure  will change continuously but the energy density
will  have a discontinuity at $ r = r_c$ . The TOV equations with
nuclear  matter equation of state are continued until the 
pressure vanishes which defines the surface of the star. This completes the 
calculations for stellar model for a hybrid  neutron star, whose mass and 
radius can be calculated for different central densities.
\begin{figure}[t]
\leavevmode
\protect\centerline{\epsfxsize=5in\epsfysize=5in\epsfbox{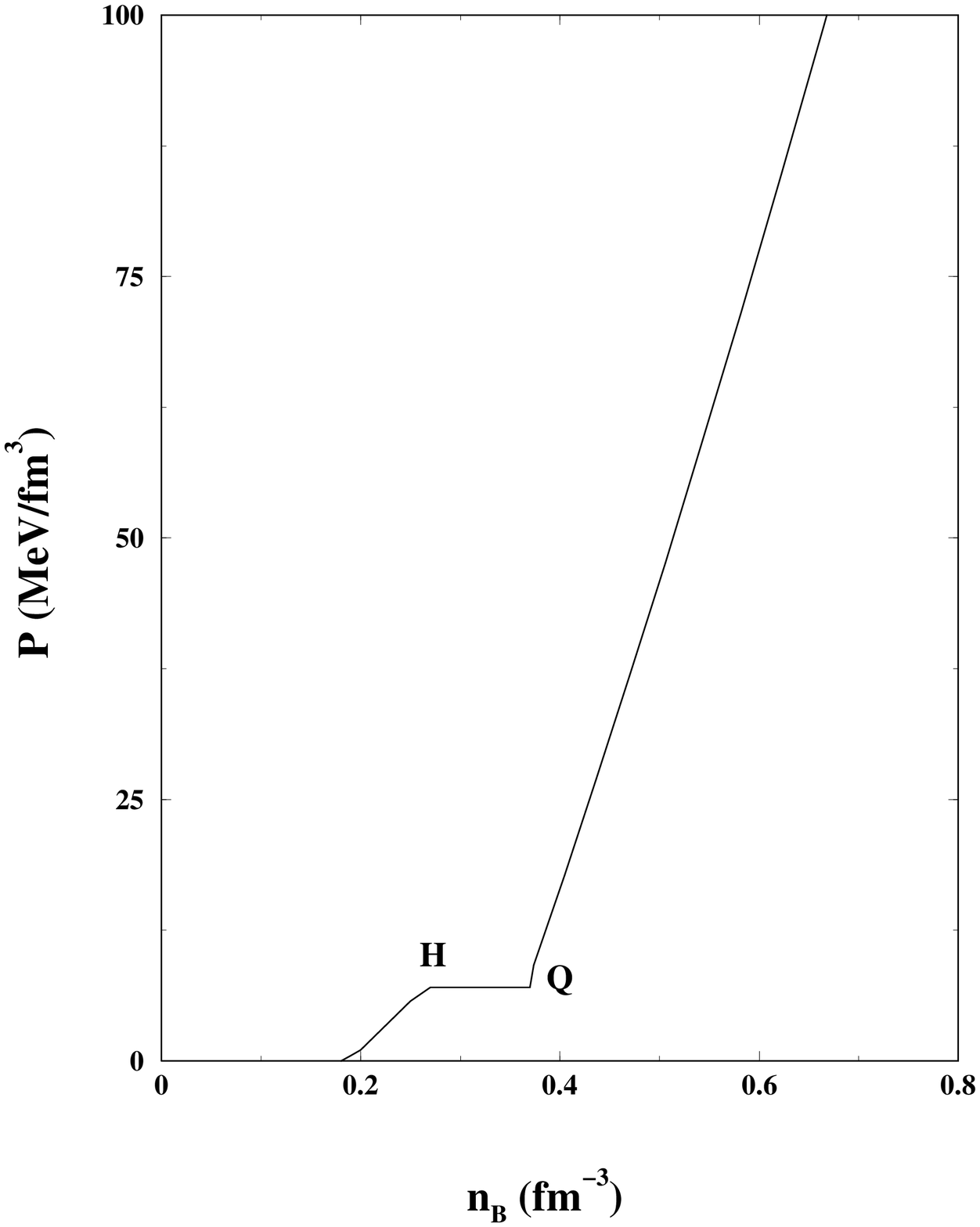}}
\caption{Pressure (P) vs. number density($n_B$) of hybrid star.} 
\end{figure}

   Fig.[4] shows behavior of pressure versus number density in the vicinity
of a first order phase transition in a system having one chemical 
potential corresponding to conserved baryon number$^{8}$. Points labeled
H and Q mark the end of the hadronic and beginning of the quark phase, the 
intervening region representing the mixed phase. The two equal pressure
points  at the opposite ends of the mixed phase are mapped onto the
same  radial point in the star. These aspects are also illustrated in
Fig.[5].
\begin{figure}[t]
\leavevmode
\protect\centerline{\epsfxsize=5in\epsfysize=5in\epsfbox{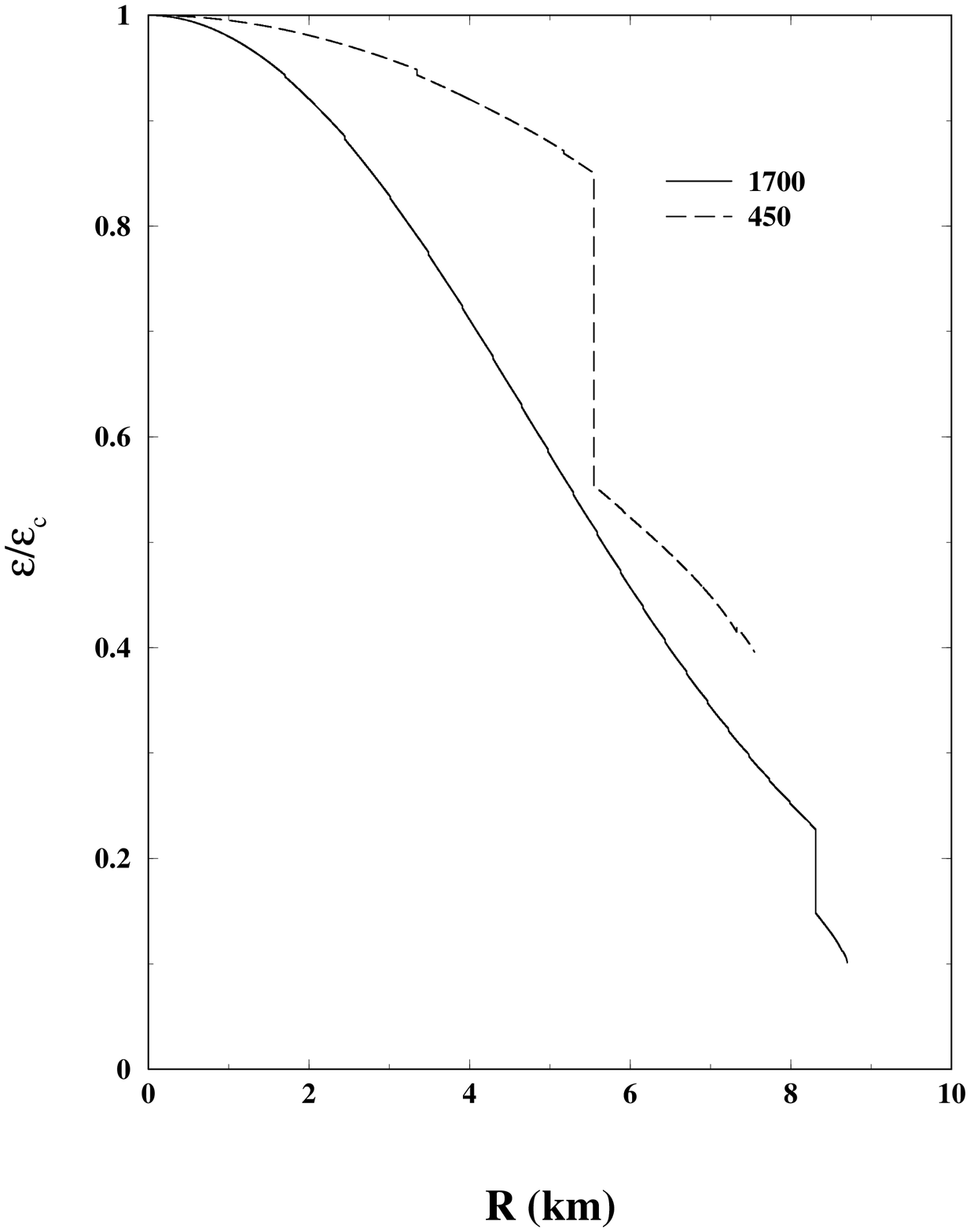}}
\caption{Energy density profile for central densities
  $\epsilon_c$ = 1700 MeV/fm$^3$(solid line)
  and$\epsilon_c$ = 450 MeV/fm$^3$(dashed line).} 
\end{figure}

  Through out the calculation of hybrid star we have used $\alpha_s$ = 
0.5 and B = (150MeV)$^4$ for quark sector.  
The energy density profile obtained from Eqns.(14-16) are plotted
in Fig.[5] for central energy density $\epsilon_{c}$= 450 MeV/fm$^3 $ and 
$\epsilon_{c}$ = 1700 MeV/fm$^3 $. For core energy densities greater than the
critical energy density ($\simeq $ 387 MeV/fm$^3$) the core consists of
quark matter. As we go away from the core towards the surface through TOV
equations, when the critical pressure is reached, the density drops 
discontinuously indicating a first order phase transition. Thus for central 
density  of 450 MeV/fm$^3$ such a star has a quark matter core of radius
5.5 kms with nuclear matter crust of about 2.2 kms , whereas for 
$\epsilon_c$ = 1700 MeV/fm$^3$, the quark matter core radius is 8.2
kms with nuclear matter crust of about 0.6 kms. Hence it is clear that 
if we take smaller central energy densities then nuclear matter is
expected to be more abundant in a hybrid star .

        We have plotted in Fig.[6] the mass  of hybrid star as a
function of  central energy density to examine the stability of such a 
star. Hence we have two branches of solutions. Pure neutron star at
lower densities $\epsilon_c < \epsilon_{nm}^{cr}$ 
and hybrid stars at central densities $\epsilon_c > \epsilon_{qm}^{cr}$.
Taking into account the stability of such  stars
under density fluctuations require dM/d$\epsilon_c > 0$$^{26}$. 
As may be seen from the figure, dM/d$\epsilon_c$ becomes negative
around 1700 MeV/fm$^3$ after which it  may
collapse into black holes$^{26,28}$. This yields the maximum mass of
hybrid star as M $\simeq$ 1.69 $M_{\odot}$. Fig.[7] shows the mass as a 
function of radius obtained for different central densities varying in the
range of 450 MeV/fm$^3$ to 1700 MeV/fm$^3$ for such a star which
indicates the maximum radius to be around 9.3 kms. To check the stability
of our result against the errors in values of symmetric energy (32$\pm$6 MeV)
and the nuclear incompressibility (300$\pm$50 MeV), we have also computed the 
coupling constants ($C_{\rho},C_{\omega},C_{\sigma}$,B and C) for four cases 
of symmetric energy value 32 MeV with incompressibility 250 and 350 MeV and
the incompressibility of 300 MeV with symmetric values 26 and 38 MeV. The 
maximum values for M and R are found to be M = 1.69$^{+0.005}_{-0.001}
M_{\odot}$ and R = 9.3$^{+0.21}_{-0.05}$ kms. Such small effect of symmetric 
energy and incompressibility on maximum values of M and R can be attributed
to the fact that only about 15$\%$ of the radius of the hybrid star contains 
nuclear matter.    
\begin{figure}[t]
\leavevmode
\protect\centerline{\epsfxsize=5in\epsfysize=5in\epsfbox{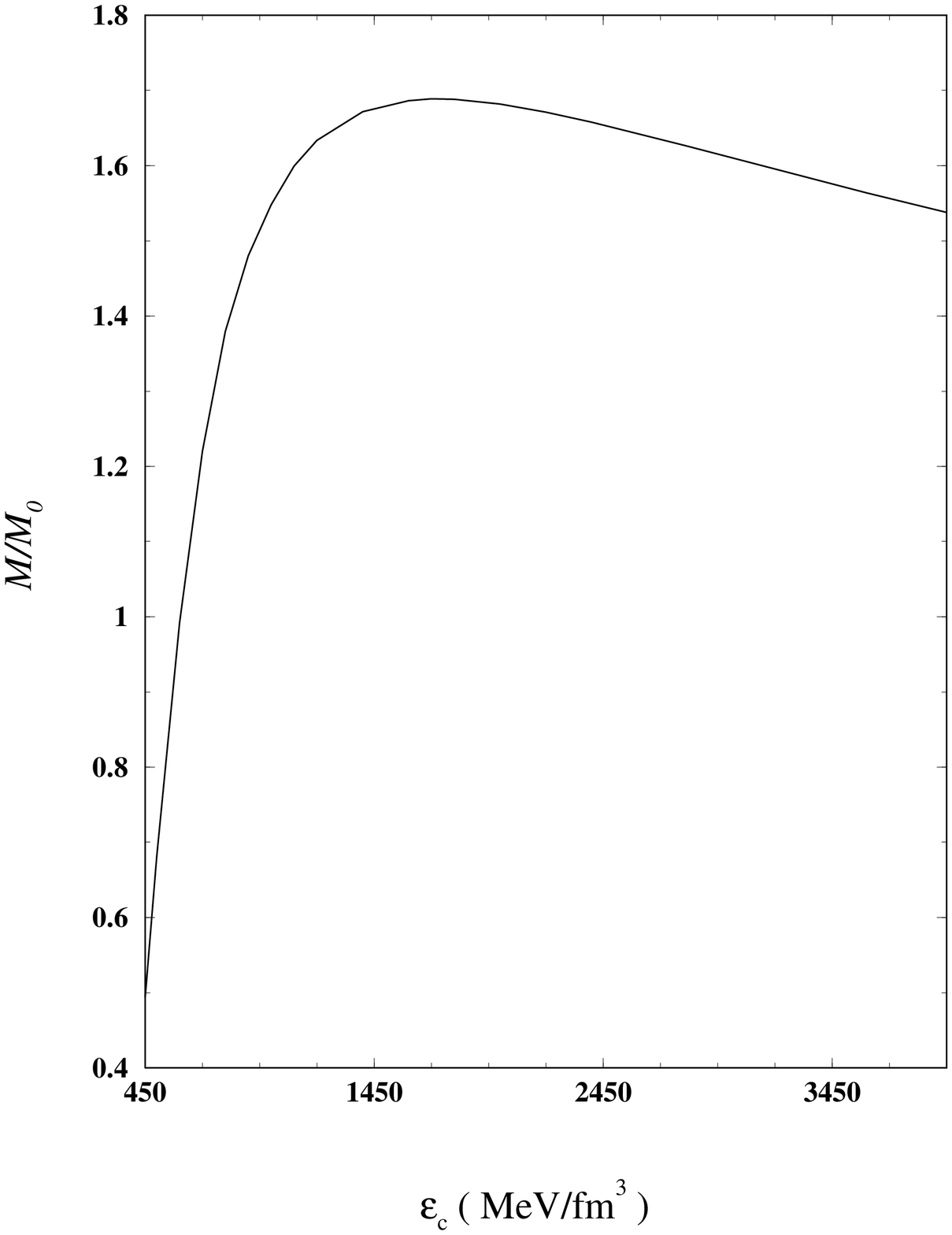}}
\caption{The mass(M/$M_{\odot}$) of the  hybrid
   star as a function of central energy density($\epsilon_c$).} 
\end{figure}

\begin{figure}[t]
\leavevmode
\protect\centerline{\epsfxsize= 5in\epsfysize=5in\epsfbox{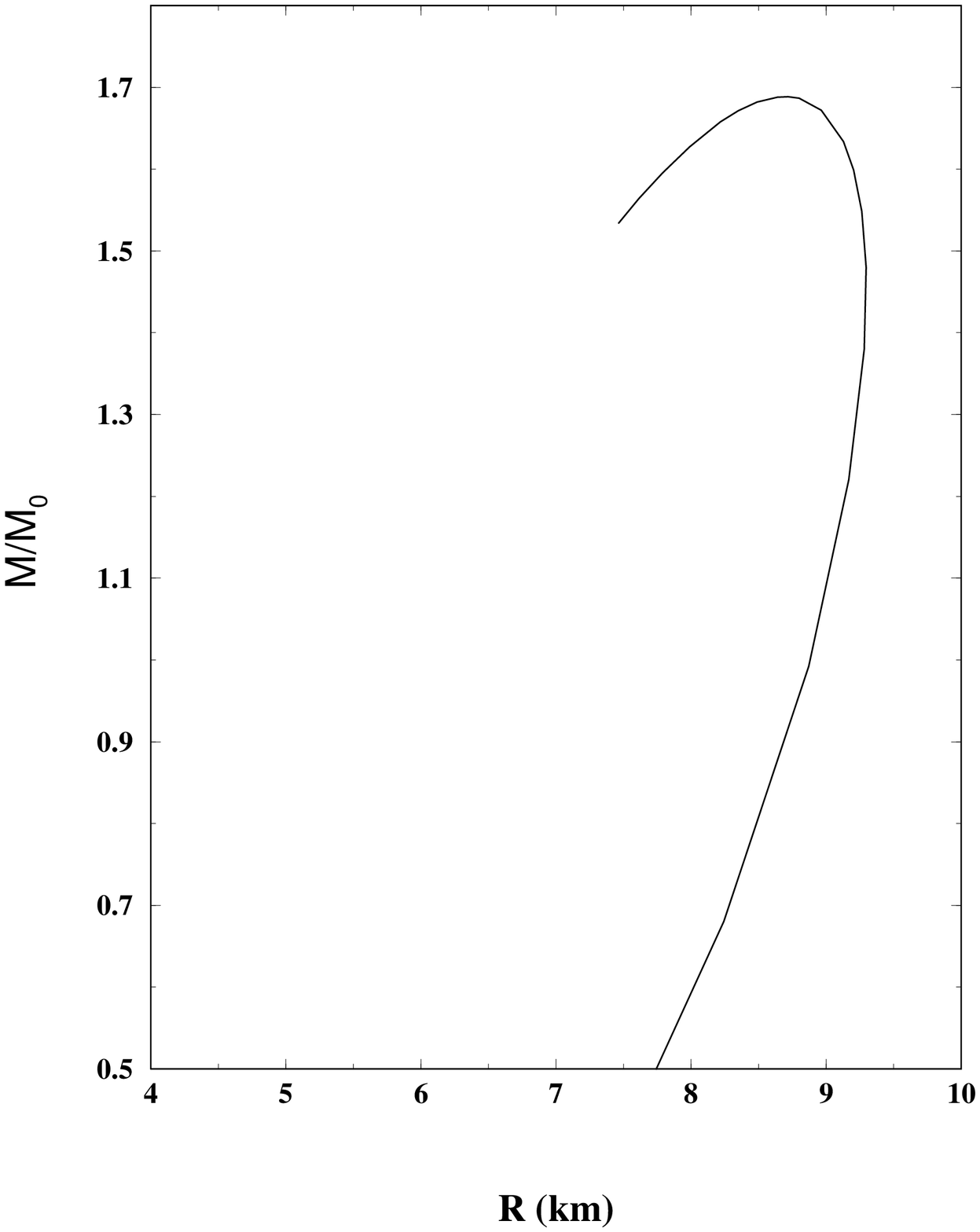}}
\caption{Mass as a function of radius for the hybrid star.} 
\end{figure}

   We also calculate the surface gravitational redshift $Z_s$ of photons 
which is given by$^{29}$

\begin{equation}
    Z_s = \frac{1}{\sqrt{(1-2GM/R)}}-1 
\end{equation}
\noindent
In Fig.[8] we have plotted $Z_s$ as a function of $M/M_{\odot}$ . 
In this context it may be mentioned here that  our result for the
surface redshifts lie in the range of 0.2 to 0.5 as 
determined from gamma ray bursters$^{30}$.

    We then compute the relativistic Keplerian angular velocity $\Omega_k$ 
given by$^{31}$

\begin{equation}
   \frac{\Omega_k}{10^4 sec^{-1}} = 0.72 \sqrt{\frac{M/M_{\odot}}
        {(R/10km)^3}}
\end{equation}
\noindent
\begin{figure}[t]
\leavevmode
\protect\centerline{\epsfxsize=5in\epsfysize=5in\epsfbox{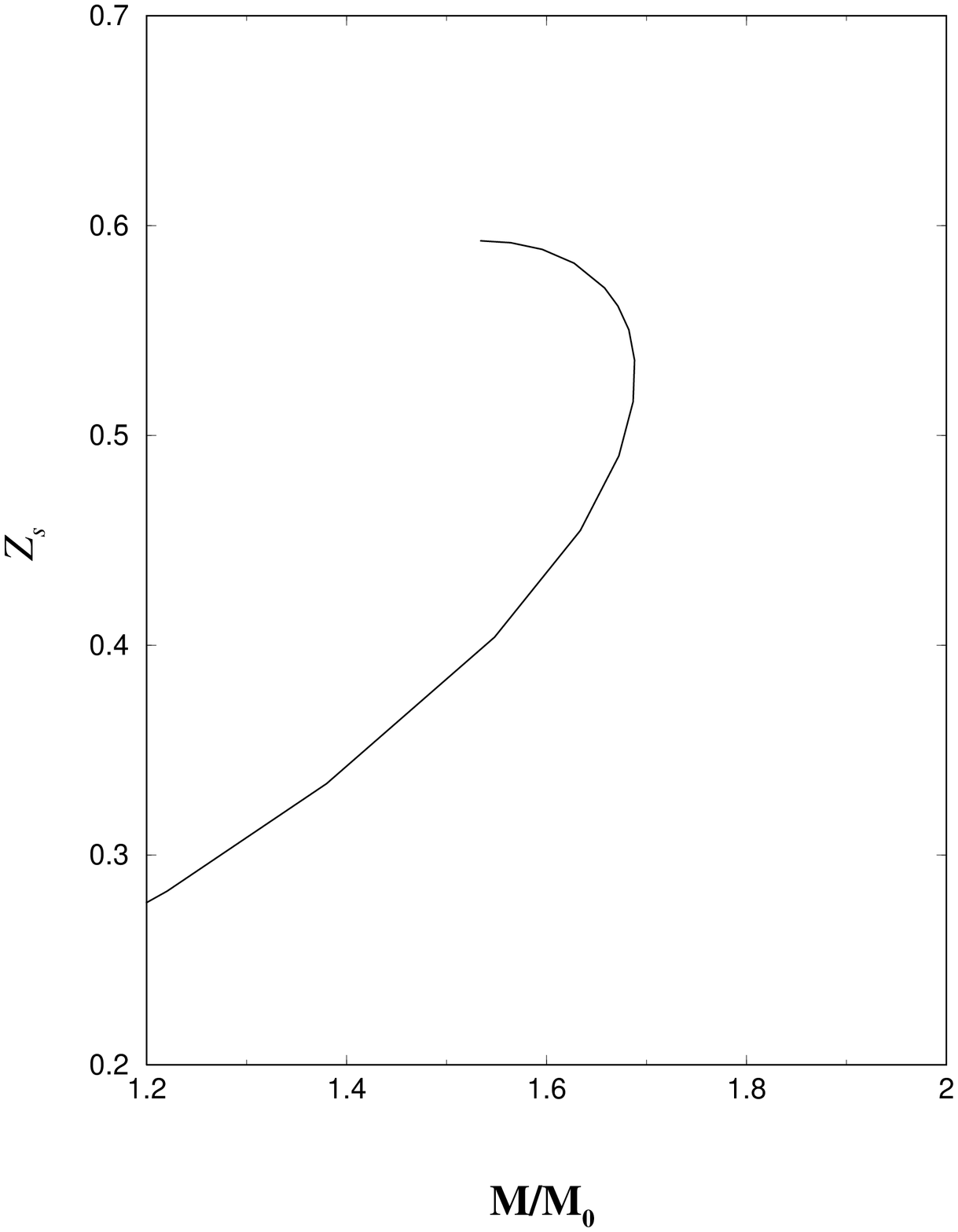}}
\caption{The surface gravitational redshift($Z_s$) as a function of star 
  mass(M/$M_{\odot}$)} 
\end{figure}
\begin{figure}[t]
\leavevmode
\protect\centerline{\epsfxsize=5in\epsfysize=5in\epsfbox{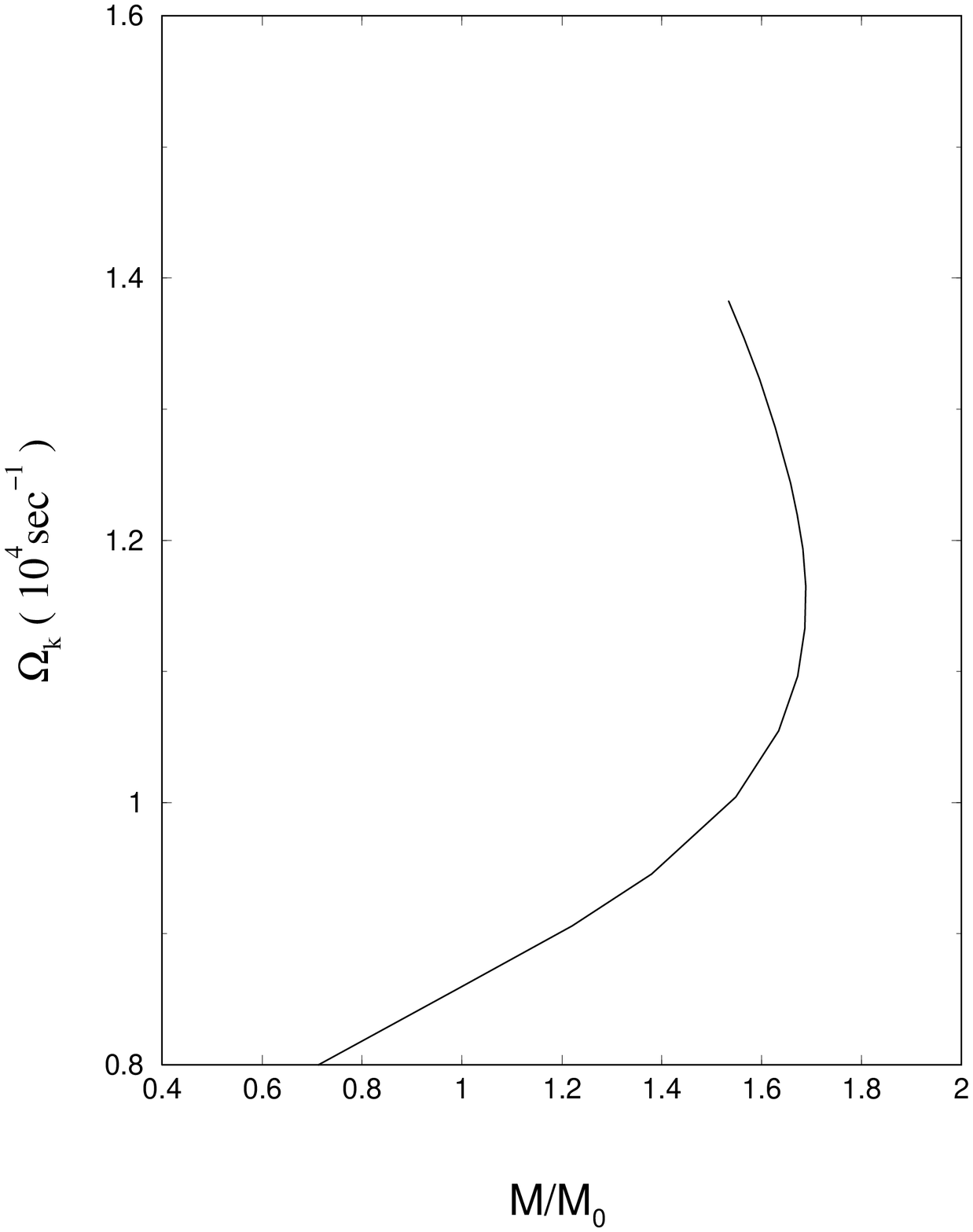}}
\caption{The Keplerian angular velocity($\Omega_k$) as a function of star 
  mass(M/$M_{\odot}$)} 
\end{figure}
\noindent
as for neutron stars. Fig.[9] shows our result for variation of relativistic 
Keplerian angular velocity as a function of $M/M_{\odot}$ for such a star.
We find that the $\Omega_k$ has an inverse relationship after an
initial increase with the mass of the star beyond 1.69$M_{\odot}$$^{4}$.
This indicates that we can not have mass of hybrid star more than  
about 1.69$M_{\odot}$. We also observe that the maximum value of
$\Omega_k$ being near about $10^4 sec^{-1}$ implies the time period to 
be less than 0.5 millisecond.


\section{\bf{Conclusions}}

  Within the formalism of a modified SU(2) chiral sigma model (Eqn.1)
we found that a first order phase transition exists between the nuclear phase 
and quark phase at density of about two to three times the nuclear matter 
density. Quark matter equation of state has the parameters: $m_s$, $\alpha_s$
and B. Our results show that the critical parameters while increasing
with increasing value of $\alpha_s$ and B, exhibit greater sensitivity
with respect to $\alpha_s$.

   The phase transition from nuclear matter to quark matter indicates that 
the core of a  neutron star consists of quark matter. To obtain a stable
hybrid star solution, we have solved TOV equations using appropriate 
equations of state with a given central energy density $\epsilon_c$  
and have taken $\alpha_s$ = 0.5 and $ B = (150 MeV)^4$
for quark matter EOS. It is observed that a stable hybrid star with a
quark core and a nuclear matter crust exist upto $\epsilon_c \simeq $
1700 MeV/fm$^3$ beyond which instability is indicated. 
For $\epsilon_c$  varying from 450 MeV/fm$^3$ to 1700 MeV/fm$^3$ the
mass of hybrid star varies from 0.5 to 1.69 $M_{\odot}$ and the radius 
from 7.7 to 8.8 kms with a quark core of about 5 to 8 kms respectively.  
Our results, thus, indicate that the bulk of hybrid star is composed of 
quark matter with a  crust of nuclear matter. 
We find that the maximum mass and radius of  hybrid stars 
to be about 1.69 $M_{\odot}$ and 9.3 kms respectively  depending on the
values of the parameters used in the model. If we compare these mass
and radius values of the hybrid stars with those obtained using this
model$^{10}$ for pure neutron star (such as $M = 2.1M_{\odot}$ and R =
12.1 kms), we find that the hybrid stars are more compact than a
normal neutron star. This result is in exact agreement with the
general expectation since the EOS of quark matter forming the core of
hybrid star is supposed to be softer than that of neutron matter
because of QCD asymptotic freedom. The greater compactness of the star 
also leads to smaller time period lying in the submillisecond range as 
obtained by us. It is also observed that
the surface garvitational redshift and relativistic Keplerian angular 
velocity of the hybrid stars can not increase beyond $M/M_{\odot}$ = 
1.69 ; showing a decrease with increase in $ M/M_{\odot}$ beyond this value. 

\vspace {0.1in}
\noindent {\bf{Acknowledgements}}
\vspace {0.05in}

   We would like to thank P.K.Sahu  for helpful discussions
and suggestions. We are also thankful to 
Institute of Physics, Bhubaneswar,
India, for providing the library and computational facility. P.K.Jena would
like to thank Council of Scientific and Industrial Research, Government of
India, for the award of JRF, F.No. 9/173 (101)/2000/EMR.We thank the Referee
for suggesting a number of improvements in the manuscript.

\newpage

\end{document}